# De mythe van geïnformeerde toestemming: online privacybescherming kan beter

Dr. Frederik Zuiderveen Borgesius

Frederikzb[at]cs.ru.nl

*De huidige privacyregels leggen veel nadruk op de geïnformeerde toestemming van internetgebruikers. Met zulke toestemmingsregels probeert de wet mensen in staat te stellen om keuzes te maken in hun eigen belang. Maar inzichten uit gedragsstudies trekken de effectiviteit van deze wetgevingstactiek in twijfel. Zo klikken internetgebruikers in de praktijk 'OK' op vrijwel elk toestemmingsverzoek dat op hun scherm verschijnt. De wet zou meer aandacht moeten geven aan het beschermen van mensen.*

**Trefwoorden:** privacy, persoonsgegevens, telecommunicatiewet, cookies, internet, gedragsstudies

**Inleiding**

Geïnformeerde toestemming speelt een hoofdrol in de huidige privacyregels. Zo mogen bedrijven alleen een *tracking cookie* plaatsen nadat de internetgebruiker daarvoor toestemming heeft gegeven. Deze bijdrage bespreekt problemen met geïnformeerde toestemming aan de hand van het voorbeeld met *behavioural targeting*. Bij deze marketingtechniek volgen bedrijven het online gedrag van mensen, en gebruiken de verzamelde informatie om hun gerichte advertenties te tonen. Bieden toestemmingsregels voldoende privacybescherming op het gebied van behavioural targeting, en zo nee: hoe zou het recht privacy beter kunnen beschermen?[1]

Paragraaf 1 en 2 bespreekt behavioural targeting en drie daaraan gerelateerde privacyproblemen. Paragraaf 3 laat zien dat geïnformeerde toestemming een hoofdrol speelt in de huidige privacyregels. Op papier lijkt deze wetgevingstechniek verstandig. Maar paragraaf 4 laat aan de hand van gedragsstudies zien dat mensen 'OK' klikken op vrijwel elk verzoek dat zij tegenkomen op het internet. Paragraaf 5 betoogt dat de wet niet alleen moet mikken op *empowerment* van het individu, maar ook op *protection*. Paragraaf 6 roept op tot een maatschappelijk debat over het gebruik van persoonlijke informatie door internetbedrijven. Paragraaf 7 bevat de conclusie: de wet zou mensen geen keuzes moeten opdringen waaraan ze niets hebben. De wet zou meer aandacht moeten hebben voor het beschermen van mensen.

**1       Internet marketing en behavioural targeting**

We kunnen veel internetdiensten, zoals websites, e-maildiensten en zoekmachines, gebruiken zonder rechtstreeks met geld te betalen. Maar via zulke diensten worden vaak wel enorme hoeveelheden gegevens over gebruikers verzameld. In een veelvoorkomend model voor online marketing betalen adverteerders een websitehouder alleen als iemand op een advertentie klikt. Het percentage mensen dat op advertenties klikt is laag: rond de 0,05% tot 0,2%.[2] De marketingtechniek behavioural targeting is ontwikkeld om dat percentage te verhogen.

Bedrijven verzamelen veel informatie over wat mensen doen op het internet: wat zij lezen, welke video's zij kijken, wat zij zoeken, etc. Onder meer Google, Facebook en

---

[1] De bijdrage is gebaseerd op het proefschrift van de auteur, getiteld 'Improving Privacy Protection in the Area of Behavioural Targeting'. Dit verschijnt in de lente van 2015 bij Kluwer Law International, in de Information Law Series. De auteur dankt Sarah Eskens, Kelly Breemen, en Gerard Mom voor commentaar op een eerdere versie van deze bijdrage.

[2] Zie bijvoorbeeld: D. Chaffey, 'Display advertising clickthrough rates', 13 November 2013 <www.smartinsights.com/internet-advertising/internet-advertising-analytics/display-advertising-clickthrough-rates/>. Alles in de voetnoten genoemde websites zijn geraadpleegd op 28 februari 2015.

Microsoft exploiteren een advertentienetwerk. Sommige bedrijven volgen het gedrag van honderden miljoenen internetgebruikers. Al deze gebruikersprofielen kunnen worden verrijkt met gegevens die on- en off-line worden verzameld, zoals locatiegegevens van telefoons.

Voor behavioural targeting worden vaak cookies of vergelijkbare technieken gebruikt. Een cookie is een klein tekstbestand dat een websitehouder op de computer van een internetgebruiker plaatst, om die computer te herkennen. Websites gebruiken cookies bijvoorbeeld voor log-in procedures, en om de inhoud van een winkelmandje te onthouden. Advertentienetwerken, bedrijven die advertenties tonen op duizenden websites, kunnen ook cookies plaatsen en lezen. Op deze manier kan een advertentienetwerk het gedrag van internetgebruikers volgen over alle websites waarop het advertenties toont.

Het is overigens omstreden of behavioural targeting nodig is om 'gratis' websites te financieren.[3] Adverteren zonder behavioural targeting is ook mogelijk. Advertenties kunnen bijvoorbeeld aansluiten bij de inhoud van een website, in plaats van bij het profiel van een internetgebruiker. Zo kunnen advertenties voor auto's geplaatst worden op websites over auto's.

## 2    Privacyproblemen van internet marketing

Het Europees Hof voor de Rechten van de Mens zegt dat informatie die wordt afgeleid uit iemands internetgebruik onder de bescherming van het recht op privacy valt.[4] De belangrijkste privacygerelateerde problemen rond behavioural targeting kunnen in drie categorieën worden ingedeeld: (i) *chilling effects*, (ii) een gebrek aan controle over persoonlijke informatie en (iii) het risico van oneerlijke discriminatie en manipulatie.[5]

---

[3] In 2011 was behavioural targeting volgens brancheorganisatie IAB goed voor twee procent van alle bestedingen aan online advertenties in Nederland. M. Hartkamp, 'Online advertising overstijgt markt voor tv-reclame', Adformatie <www.adformatie.nl/nieuws/bericht/online-advertising-overstijgt-markt-voor-tv-reclame>.
[4] EHRM, Copland v. United Kingdom, No. 62617/00, 3 april 2007, par. 41.
[5] Zie voor een vergelijkbare indeling: B. van der Sloot, 'Het plaatsen van cookies ten behoeve van behavioural targeting vanuit privacyperspectief', *Privacy & Informatie*, 2011-2, 62.

Ten eerste kan de grootschalige gegevensverzameling voor behavioural targeting een *chilling effect* veroorzaken. Mensen passen zich aan als zij weten dat hun gedrag wordt geobserveerd. Een bezoek aan een website over een ziekte of een politieke partij voelt ongemakkelijk als je weet dat je bezoek mogelijk geregistreerd wordt door allerlei partijen.[6]

Ten tweede hebben mensen amper controle over de hen betreffende gegevens. Massale gegevensopslag brengt risico's met zich mee. Zo kan een datalek optreden, of kunnen gegevens worden gebruikt voor onverwachte doeleinden, zoals identiteitsfraude. Ook is het gevoel geen controle te hebben over de eigen gegevens op zichzelf al een privacyprobleem.[7]

Ten derde maakt behavioural targeting discriminatie mogelijk. Bedrijven kunnen mensen indelen in 'targets' en 'waste', en hen zo behandelen.[8] Een adverteerder kan kortingen gebruiken om welvarende mensen te verleiden om klant te worden – en tegelijkertijd armere mensen uitsluiten van een campagne. Sommigen vrezen dat gepersonaliseerde reclame zo effectief kan worden dat adverteerders een oneerlijk voordeel verkrijgen ten opzichte van het individu.[9] Ook wordt gewezen op het risico van *filter bubbles*: als algoritmes bepalen wat we zien op het web, ziet ieder van ons een ander beeld van de wereld.[10] Op die manier zouden gepersonaliseerde diensten ongemerkt ons gedrag kunnen beïnvloeden. Het risico op filter bubbles lijkt vooral relevant als niet alleen advertenties, maar ook websites, zoekmachines of andere diensten worden gepersonaliseerd.

---

[6] Zie N.M. Richards, 'Intellectual privacy', *Texas Law Review* (2008) 87, 387.
[7] Zie S. Gürses, *Multilateral Privacy Requirements Analysis in Online Social Networks (diss. Universiteit Leuven)*, 2010, p 87-91; MR Calo, 'The boundaries of privacy harm', *Indiana Law Journal* (2011) 86, 1131.
[8] Zie J. Turow, *The Daily You: How the New Advertising Industry is Defining Your Identity and Your Worth,* New Haven: Yale University Press 2011.
[9] Zie M.R. Calo, 'Digital market manipulation', *George Washington Law Review* 995 (2014) 82.
[10] E. Pariser, *The Filter Bubble*, London: Penguin Viking 2011, p 9. Zie ook M Oostveen, 'World Wide Web of Your Wide Web? Juridische aspecten van zoekmachinepersonalisatie', *Tijdschrift voor Internetrecht* (2012) 6.

## 3 Geïnformeerde toestemming in privacyrecht

Geïnformeerde toestemming speelt een centrale rol in de Europese privacyregels.[11] Het recht op bescherming van persoonsgegevens in het Handvest van de Grondrechten van de Europese Unie zegt bijvoorbeeld: "Deze gegevens moeten eerlijk worden verwerkt, voor bepaalde doeleinden en met toestemming van de betrokkene of op basis van een andere gerechtvaardigde grondslag waarin de wet voorziet."[12] Verder volgt sinds 2009 uit de e-Privacyrichtlijn, kort gezegd, dat cookies alleen geplaatst mogen worden als de internetgebruiker daarvoor geïnformeerde toestemming heeft gegeven.[13] Er zijn uitzonderingen op het toestemmingsvereiste: er is bijvoorbeeld geen toestemming nodig voor het plaatsen van cookies die worden gebruikt voor een digitaal winkelwagentje, of voor log-in procedures.

De wettelijke definitie van toestemming eist voor geldige toestemming een vrijwillige, specifieke, op informatie berustende wilsuiting.[14] Wilsuitingen zijn doorgaans niet aan vormvereisten onderworpen, maar een wilsuiting kan vrijwel nooit afgeleid worden uit niets doen.

Sommige marketeers suggereren dat iemand impliciet toestemming geeft voor behavioural targeting als hij tracking cookies niet blokkeert in zijn browser.[15] Deze interpretatie van de regels overtuigt niet. Als iemand de standaardinstellingen van zijn browser niet aanpast, uit hij daarmee nog niet zijn wil om tracking cookies toe te staan waarmee zijn gedrag wordt gevolgd.[16]

---

[11] Het recht op bescherming van persoonsgegevens (artikel 8 van het Handvest) moet onderscheiden worden van het recht op privacy (artikel 7). Ter wille van de leesbaarheid spreekt deze bijdrage over privacyrecht in plaats van persoonsgegevensbeschermingsrecht. Zie over het verschil: G. González Fuster, *The Emergence of Personal Data Protection as a Fundamental Right of the EU* (diss. Vrije Universiteit Brussel), Dordrecht: Springer 2014.
[12] Artikel 8(2) van het Handvest van de Grondrechten van de Europese Unie.
[13] Artikel 5(3) van de e-Privacyrichtlijn (Richtlijn 2002/58/EG, laatst gewijzigd door Richtlijn 2009/136/EG). Artikel 5(3) is in Nederland geïmplementeerd in de Telecommunicatiewet (artikel 11.7a).
[14] Artikel 2(f) van de e-Privacyrichtlijn verwijst voor de definitie van toestemming naar artikel 2(h) van de Richtlijn bescherming persoonsgegevens (Richtlijn 95/46/EG).
[15] Het Interactive Advertising Bureau van het Verenigd Koninkrijk, een brancheorganisatie, zegt bijvoorbeeld: 'We believe that default web browser settings can amount to 'consent' (…)'. (Interactive Advertising Bureau United Kingdom, 'Department for Business, Innovation & Skills consultation on implementing the revised EU electronic communications framework, IAB UK Response' (1 December 2012) <www.iabuk.net/sites/default/files/IABUKresponsetoBISconsultationonimplementingtherevisedEUElectronicCommunicationsFramework_7427_0.pdf>).

Toestemming moet 'vrijwillig' worden gegeven. Maar de Europese privacyregels verbieden bedrijven niet expliciet om een *take-it-or-leave-it* keuze te bieden. In veel gevallen is het niet verboden voor websitehouders om een trackingmuur te installeren, en zo toegang te weigeren aan bezoekers die geen toestemming geven voor tracking cookies.[17] Maar als mensen een website *moeten* gebruiken dwingt een trackingmuur toestemming af, en is de toestemming dus ongeldig.[18]

## 4    Geïnformeerde toestemming in de praktijk

Met toestemmingsregels probeert het privacyrecht mensen in staat te stellen om keuzes te maken in hun eigen belang. Maar inzichten uit gedragsstudies trekken de effectiviteit van deze wetgevingstactiek in twijfel.

Er is sprake van informatieasymmetrie: internetgebruikers weten nauwelijks dat hun gegevens verzameld worden, waarvoor die gebruikt worden, en wat de consequenties kunnen zijn.[19] Voor websitebezoekers is het bijvoorbeeld moeilijk te zien of hun gegevens worden verzameld en wat daarmee gebeurt. Websitehouders hebben dan ook weinig prikkels om af te zien van behavioural targeting. Bovendien: als alle concurrenten profiteren van informatieasymmetrie, kan een bedrijf uit de markt gedrukt worden als het niet hetzelfde doet.[20] Inderdaad concurreren websites zelden op privacy.

---

[16] Artikel 29 Werkgroep, 'Opinion 2/2010 on online behavioural advertising' (WP 171), 22 juni 2010.
[17] European Agency for Fundamental Rights, *Handbook on European data protection law,* Brussel: Publications Office of the European Union 2014, p. 59.
[18] De Artikel 29 Werkgroep, een samenwerkingsverband van Europese privacytoezichthouders, benadrukt dat toestemming vrijwillig moet zijn, maar zegt niet dat de wet trackingmuren onder alle omstandigheden verbiedt. Artikel 29 Werkgroep, 'Working Document 02/2013 providing guidance on obtaining consent for cookies' (WP 208) 2 oktober 2013.
[19] A. Acquisti & J. Grossklags , 'What Can Behavioral Economics Teach Us About Privacy?' in: A. Acquisti e.a. (red.), *Digital Privacy: Theory, Technologies and Practices,* Londen: Taylor and Francis Group 2007.
[20] T. Vila, R. Greenstadt & D. Molnar, 'Why We Can't be Bothered to Read Privacy Policies. Models of Privacy Economics as a Lemons Market' in: L.J. Camp en S. Lewis (red.), *Economics of Information Security,* Heidelberg: Springer 2004.

Zo worden tracking cookies geplaatst via vrijwel elke populaire website.[21] Voor smartphone applicaties is de situatie vergelijkbaar.

De Europese privacyregels proberen de informatieasymmetrie te verkleinen: bedrijven moeten de betrokkene alle informatie geven die noodzakelijk is om een eerlijke verwerking van persoonsgegevens te waarborgen.[22] Websitehouders kunnen een privacyverklaring gebruiken om te voldoen aan die transparantieverplichtingen.[23]

Maar bij behavioural targeting is het de vraag of een websitehouder bezoekers wel kan uitleggen wat er met hun gegevens gebeurt. Vaak worden bij één websitebezoek cookies geplaatst door tientallen bedrijven, en die bedrijven verkopen de gegevens soms weer door. Volgens een grote uitgever, die tracking cookies laat plaatsen via haar websites, is het 'praktisch onmogelijk om honderd procent zeker te weten wat derde partijen precies van je site halen.'[24] Een ander probleem is dat privacyverklaringen vaak ingewikkeld, vaag en lang zijn. Er is berekend dat het mensen enkele weken per jaar zou kosten als zij de privacyverklaring van elke bezochte website zouden lezen.[25] Vrijwel niemand leest dan ook privacyverklaringen.[26] In de praktijk lost het privacyrecht de informatieasymmetrie dus niet op.

Bovendien blijkt uit gedragsstudies dat ons doen en laten wordt beïnvloed door verschillende *biases*, zoals de *default bias* en de *present bias*. De default bias beschrijft de neiging om geen actieve keuze te maken.[27] In een opt-out regime, waarin mensen worden verondersteld in te stemmen als zij geen bezwaar maken, geven mensen al snel

---

[21] C.J. Hoofnagle & N. Good, 'The web privacy census', oktober 2012 <http://law.berkeley.edu/privacycensus.htm>.

[22] Artikel 10 en 11 van de Richtlijn bescherming persoonsgegevens. De richtlijn legt verplichtingen op aan 'verantwoordelijken' (artikel 2(d)). Ter wille van de leesbaarheid spreekt deze bijdrage van bedrijven.

[23] E.W. Verhelst, *Recht Doen aan Privacyverklaringen: een Juridische Analyse van Privacyverklaringen op Internet* (diss. Universiteit van Tilburg), Verhelst 2012.

[24] De Telegraaf Media Group, geciteerd in M. Martijn, 'Big Business is watching you', De Correspondent, 9 oktober 2013 <https://decorrespondent.nl/66/Big-Business-is-watching-you/3214002-df572412>.

[25] AM McDonald & LF Cranor, 'The Cost of Reading Privacy Policies', *I/S: A Journal of Law and Policy for the Information Society* (2008) 4(3) 540.

[26] F. Marotta-Wurgler, 'Will Increased Disclosure Help? Evaluating the Recommendations of the ALI's Principles of the Law of Software Contracts' *The University of Chicago Law Review* (2011) 78(1) 165

[27] W. Samuelson & R. Zeckhauser, 'Status Quo Bias in Decision Making' *Journal of Risk and Uncertainty* (1988) 1(1) 7.

'toestemming'. Daarentegen geven mensen minder snel toestemming in een opt-in regime dat een wilsuiting vereist.[28]

Present bias beschrijft de neiging om nú voordeel te kiezen, en nadeel in de toekomst te negeren. Veel mensen vinden het bijvoorbeeld moeilijk om geld te sparen of een dieet vol te houden.[29] De present bias heeft ook invloed op keuzes over privacy. Als websitebezoekers op een trackingmuur stuiten, zullen zij tracking cookies waarschijnlijk accepteren, en de nadelen van toekomstige privacy-inbreuken negeren.[30]

Kortom, de praktische problemen met geïnformeerde toestemming lijken vrijwel onoplosbaar. Privacyverklaringen en toestemmingsverzoeken worden niet gelezen; ze zouden toch niet worden begrepen; en als ze begrepen zouden worden, dan zouden mensen daar niet naar handelen.

## 5    Protection en empowerment

Gezien de beperkte mogelijkheden van geïnformeerde toestemming als privacybeschermingsmaatregel, zou de wet niet alleen moeten streven naar *empowerment* van het individu, maar ook naar *protection*.

Empowerment betreft maatregelen om mensen in staat te stellen om voor hun eigen belangen op te komen. Streven naar empowerment volstaat niet om privacy te beschermen in de context van behavioural targeting. Maar enige verbetering moet mogelijk zijn, in vergelijking met de huidige situatie waarbij het gedrag van miljoenen mensen heimelijk wordt gevolgd. Om de informatieasymmetrie te verminderen, zou de wetgever moeten afdwingen dat toestemmingsverzoeken simpel, kort en begrijpelijk zijn. Verder moeten bestaande toestemmingsregels beter gehandhaafd worden. De 'wie zwijgt stemt toe'-redenering van veel websites zou niet geaccepteerd moeten worden.

---

[28] Zie A. Acquisti A & R Gross, 'Imagined Communities: Awareness, Information Sharing, and Privacy on the Facebook' (2006) 4258 6th International Workshop, PET 2006, Cambridge, UK, 28-30 juni, 2006 (Lecture Notes in Computer Science) 36.
[29] C.R. Sunstein & R.H. Thaler, *Nudge: Improving Decisions about Health, Wealth, and Happiness*, Newhaven: Yale University Press 2008.
[30] Uit een enquête door de Consumentenbond uit 2014 bleek: 'van de 50% die altijd op 'OK' klikt, wil 32% van geen enkele site tracking cookies' (Consumentenbond, 'Cookiewet heeft bar weinig opgeleverd' <www.consumentenbond.nl/test/elektronica-communicatie/veilig-online/privacy-op-internet/extra/cookiewet-heeft-weinig-opgeleverd/>).

Een wettelijk privacyregime zonder enige rol voor geïnformeerde toestemming is ook niet goed voorstelbaar. Het is niet haalbaar om alle wenselijke en onwenselijke gegevensverwerkingen vooraf te definiëren. Bovendien: smaken verschillen – ook op het gebied van privacy. Wat de één toestaat, zou een ander weigeren. Kortom, in veel gevallen zal een toestemmingsregel, in combinatie met de waarborgen uit het privacyrecht, waarschijnlijk de juiste aanpak blijven. In die gevallen moeten transparantie en toestemming serieus genomen worden.

De tweede juridische tactiek voor privacybescherming ziet meer op *protection* dan op empowerment. Verschillende regels uit de Europese Richtlijn bescherming persoonsgegevens kunnen privacy beschermen.[31] Als de richtlijn volledig nageleefd zou worden, dan zou zij een redelijke mate van privacybescherming bieden. De richtlijn eist bijvoorbeeld passende beveiliging voor opgeslagen persoonsgegevens.[32] Ook volgt uit de richtlijn dat gegevens die verzameld zijn voor één doel, niet zomaar voor een ander doel gebruikt mogen worden.[33] De richtlijn bevat dwingend recht, en geldt ook nadat iemand toestemming heeft gegeven voor de verwerking van zijn persoonsgegevens.[34]

De richtlijn reguleert het verwerken van persoonsgegevens: gegevens over een geïdentificeerde of identificeerbare persoon.[35] De richtlijn kan alleen helpen privacy te beschermen als zij van toepassing is. Daarom is allereerst nodig dat de gegevens die worden verwerkt voor behavioural targeting doorgaans als 'persoonsgegevens' worden gekwalificeerd.

De Artikel 29 Werkgroep, een samenwerkingsverband van Europese privacytoezichthouders, zegt dat behavioural targeting doorgaans de verwerking van persoonsgegevens met zich meebrengt – ook als een bedrijf geen persoonsnaam koppelt aan de gegevens. Als een bedrijf gegevens wil gebruiken om een persoon te onderscheiden binnen een groep, dan zijn die gegevens persoonsgegevens volgens de Werkgroep.[36] Nationale privacytoezichthouders volgen vaak de interpretatie van de Werkgroep.[37] De documenten van de Werkgroep zijn invloedrijk, maar niet juridisch bindend.

---

[31] Richtlijn bescherming persoonsgegevens, 96/46/EG.
[32] Artikel 17 van de Richtlijn bescherming persoonsgegevens.
[33] Artikel 6(1)(b) van de Richtlijn bescherming persoonsgegevens.
[34] HvJEU, Zaak C-131/12, Google Spain SL en Google Inc. tegen Agencia Española de Protección de Datos & Mario Costeja González, nog niet gepubliceerd, par 71.
[35] Artikel 2(a) van de Richtlijn bescherming persoonsgegevens.

De richtlijn kent strengere regels voor 'bijzondere categorieën gegevens', zoals persoonsgegevens omtrent ras, politieke overtuiging, gezondheid of seksleven.[38] In Nederland is het gebruik van zulke bijzondere gegevens voor behavioural targeting en andere vormen van direct marketing alleen toegestaan na *uitdrukkelijke* toestemming van de betrokkene.[39]

Strikte handhaving van de regels voor bijzondere gegevens zou privacyproblemen zoals *chilling effects* verminderen. De context van de gegevensverzameling zou in aanmerking genomen kunnen worden om te bepalen of er sprake is van bijzondere gegevens.[40] Advertentienetwerken zouden bijvoorbeeld kunnen afleiden dat iemand die websites over kanker bezoekt aan die ziekte lijdt. Als gegevens die verzameld worden via zo'n website met medische informatie als bijzondere gegevens worden gekwalificeerd, zouden die alleen verzameld mogen worden na uitdrukkelijke toestemming van de betrokkene.

---

[36] Artikel 29 Werkgroep, 'Opinion 2/2010 on online behavioural advertising' (WP 171), 22 June 2010, p. 9; Artikel 29 Werkgroep, 'Opinion 05/2014 on Anonymisation Techniques' (WP 216) 10 April 2014. In de Nederlandse Telecommunicatiewet is een rechtsvermoeden opgenomen: behavioural targeting wordt vermoed de verwerking van persoonsgegevens met zich mee te brengen (artikel 11.7a (1)). De uitleg die de Werkgroep geeft aan het begrip persoonsgegevens wordt overigens niet door iedereen onderschreven. Zie voor kritiek op de Werkgroep: G.J. Zwenne, *De verwaterde privacywet*, oratie 12 april 2013 (Universiteit Leiden). Instemmend met de Werkgroep: R. Leenes, 'Do they know me? Deconstructing identifiability' *University of Ottawa Law and Technology Journal* (2008) 4(1-2) 135.
[37] Zie over de Artikel 29 Werkgroep: Gutwirth S & Poullet Y, 'The contribution of the Article 29 Working Party to the construction of a harmonised European data protection system: an illustration of 'reflexive governance'?' in: V.P. Asinari & P. Palazzi (red.), *Défis du Droit à la Protection de la Vie Privée. Challenges of Privacy and Data Protection Law*, Bruylant 2008.
[38] Artikel 8, lid 1, Richtlijn bescherming persoonsgegevens.
[39] Artikel 23, lid 1, sub a, van de Wet bescherming persoonsgegevens.
[40] Zie over de relevantie voor de context voor privacyvraagstukken: H. Nissenbaum, *Privacy in context: technology, policy, and the integrity of social life*, Stanford: Stanford Law Books 2010.

Omdat de privacyrisico's van het gebruik van gezondheidsgegevens voor behavioural targeting zwaarder wegen dan de maatschappelijke voordelen van dergelijke praktijken, zou de wetgever een verbod op zulke praktijken moeten overwegen – of de betrokkene toestemming geeft of niet.

## 6    Oproep tot een maatschappelijk debat

Het wordt tijd dat we de privacydiscussie verbreden. Welk gebruik van persoonsgegevens vinden we aanvaardbaar in onze maatschappij, en welk gebruik niet?[41] Is het bijvoorbeeld acceptabel dat vrijwel alle nieuwsbronnen op het internet het leesgedrag van gebruikers in kaart brengen? Is het aanvaardbaar dat websites voor kinderen tracking cookies plaatsen? Is het acceptabel als webwinkels prijzen aanpassen aan het cookie-profiel van een internetgebruiker? Misschien moeten sommige praktijken gewoon verboden worden.

Het zou niet vreemd zijn om grenzen te stellen in de wet. Om consumenten te beschermen, gebruikt het recht bijvoorbeeld een combinatie van empowerment en protection. In veel gevallen verplicht de wet bedrijven om consumenten informatie te verschaffen, zodat consumenten goede keuzes kunnen maken (empowerment).[42] Maar in andere gevallen beschermt de wet consumenten rechtstreeks (protection). Op het gebied van voedselveiligheid gelden bijvoorbeeld strenge eisen,[43] en voor veel producten gelden minimum veiligheidsvoorschriften.[44]

De wet zou bijvoorbeeld in bepaalde omstandigheden tracking-muren en vergelijkbare *take-it-or-leave-it* keuzes kunnen verbieden. Zo is in Nederland onlangs de Telecommunicatiewet aangepast. Voortaan mogen websites van krachtens publiekrecht ingestelde rechtspersonen, zoals de publieke omroep, geen trackingmuur gebruiken.[45] Los daarvan is het de vraag of het überhaupt gepast is voor overheidswebsites om bedrijven in staat te stellen mensen te volgen voor behavioural targeting. De wetgever zou een verbod hierop moeten overwegen.

---

[41] Zie voor een interessante aftrap voor zo'n debat: B. Van der Sloot, 'Privacy in het post NSA-tijdperk. Tijd voor een fundamentele herziening?', *NJB* 2014-17, 1172. Zie ook L. Moerel, 'Big Data protection. How to make the draft EU Regulation on Data Protection future proof' (oratie Universiteit van Tilburg, 14 februari 2014).
[42] Zie bijvoorbeeld de Richtlijn Consumentenrechten 2011/83/EU.
[43] B. van der Meulen & M. van der Velde, *Food Safety Law in the European Union,* Wageningen: Academia Publishers 2004.
[44] Zie bijvoorbeeld de Richtlijn 2001/95/EG inzake algemene productveiligheid.

Wellicht zijn specifieke regels nodig om privacy te beschermen in de context van behavioural targeting. Zulke specifieke regels kunnen in een aparte wet worden opgenomen, maar ook elders, zoals in het mediarecht of consumentenrecht.

# 7   Conclusie

Er is geen wondermiddel voor privacybescherming op het internet. Zoals het Europees Hof voor de Rechten van de Mens opmerkt, moet privacybescherming niet theoretisch en illusoir zijn, maar praktisch en effectief.[46] Terwijl de huidige regelgeving veel nadruk legt op *empowerment* van het individu, pleit deze bijdrage voor een gecombineerde aanpak van *protection* en *empowerment*. Maar de beperkte mogelijkheden van geïnformeerde toestemming als privacybeschermingsmaatregel kunnen niet genegeerd worden. Daarom moet de wetgever relatief meer aandacht geven aan *protection*. Als de maatschappij beter af is als bepaalde praktijken niet plaatsvinden, moet een verbod overwogen worden. Het is moeilijk om de juiste balans te vinden tussen privacybescherming en ongebreideld paternalisme. Maar de wetgever moet moeilijke keuzes niet uit de weg gaan.

\* \* \*

---

[45] Art. 11.7a, lid 5. Wijziging van de Telecommunicatiewet (wijziging artikel 11.7a), gewijzigd voorstel van wet, 7 oktober 2014, Eerste Kamer, vergaderjaar 2014–2015, 33 902, A.
[46] EHRM, Christine Goodwin v UK (nr. 28957/95) (2002) 35 EHRR 18, par. 74.